# Proposal for a GHz Count Rate Near-IR Single-Photon Detector Based on a Nanoscale Superconducting Transition Edge Sensor

Daniel F. Santavicca[a], Faustin W. Carter[b], Daniel E. Prober[a,b,*]
[a]Department of Applied Physics, Yale University, New Haven, CT 06520
[b]Department of Physics, Yale University, New Haven, CT 06520

## ABSTRACT

We describe a superconducting transition edge sensor based on a nanoscale niobium detector element. This device is predicted to be capable of energy-resolved near-IR single-photon detection with a GHz count rate. The increased speed and sensitivity of this device compared to traditional transition edge sensors result from the very small electronic heat capacity of the nanoscale detector element. In the present work, we calculate the predicted thermal response time and energy resolution. We also discuss approaches for achieving efficient optical coupling to the sub-wavelength detector element using a resonant near-IR antenna.

## 1. INTRODUCTION

Near-IR single-photon detection has important applications in quantum information,[1] photon-starved classical communication,[2] spectroscopy,[3] and CMOS circuit failure analysis.[4] Figures of merit for a single-photon detector include the detection efficiency, the maximum count rate, the dark count rate, and the jitter (the uncertainty in the photon arrival time). Single-photon detectors can be either energy-resolving, with a response proportional to the photon energy, or non-energy-resolving, with a response independent of the photon energy.

Commercially available near-IR single-photon detectors include the photomultiplier tube (PMT)[5] and the single-photon avalanche diode (SPAD).[6] These detectors are typically cooled to ≈ 200 K for optimal sensitivity. Both are non-energy-resolving, although they can be operated in a photon number-resolving mode.[1] High sensitivity near-IR single-photon detection has also been demonstrated with superconducting detectors, including the transition edge sensor (TES)[7,8] and the superconducting nanowire single-photon detector (SNSPD).[9,10] The TES is inherently energy-resolving, while the SNSPD is not. These superconducting devices require significantly lower operating temperatures than the PMT and the SPAD.

In table 1 we compare several key figures of merit for these different detector technologies for 1550 nm (0.80 eV) detection. The TES offers energy resolution and the highest detection efficiency, but it also requires the lowest operating temperature (< 0.1 K) and has the slowest response time (~ 1 μs).[7] In the present work, we propose a new device, the niobium (Nb) nano-TES, that combines the energy resolution of the conventional TES with a significantly faster response time (≈ 40 ps) and a more convenient operating temperature (4 K). The proposed device is based on a superconducting Nb detector element that is 200 nm long, 100 nm wide, and 12 nm thick. Leads for device biasing and readout are made from a high conductivity metal such as gold or aluminum. The superconducting critical temperature ($T_c$) for this thickness of Nb is ≈ 5 K. The predicted values for the response time and the energy resolution of the Nb nano-TES are given in table 1.

Key to this device concept is the fact that both the energy resolution and the thermal response time can be reduced by reducing the volume, and hence the electronic heat capacity, of the Nb element. Related work has studied a titanium (Ti) nano-TES with Nb contacts.[11,12] In that device, the larger superconducting energy gap in the Nb confines hot electrons in the Ti. This hot electron confinement minimizes the thermal conductance, which improves the sensitivity for a device operated as a power detector. For single-photon (energy) detection, however, the sensitivity depends not on the thermal conductance but on the heat capacity. In the hot electron regime, the relevant heat capacity is the electronic heat capacity

*daniel.prober@yale.edu; www.yale.edu/proberlab

$C_e$. Minimizing $C_e$ by reducing the active device volume not only increases the sensitivity of a single-photon detector, it also decreases the thermal response time, resulting in a detector that is both fast and sensitive. The calculations of the predicted response time and sensitivity are described in sections 2 and 3.

Table 1. Comparison of 1550 nm single-photon detectors. Values are experimental except for the Nb nano-TES, for which predicted values are given.

| Detector | Operating temp. (K) | Detection efficiency | Jitter | Dark count rate | Response time | Energy resolving? ($\delta E_{FWHM}$) |
|---|---|---|---|---|---|---|
| PMT[5] | 200 K | 2 % | 400 ps | 200 kHz | ~ ns | No |
| SPAD[6] | 200 K | 20 % | 300 ps | 7 kHz | ~ ns | No |
| SNSPD[9] | 2.7 K | 24 % | 30 ps[10] | 1 kHz | 10 ns | No |
| Tungsten TES[7] | 0.07 K | 95 % | - | - | 0.8 μs | Yes (0.29 eV) |
| Nb nano-TES (this work) | 4 K | - | - | - | 40 ps | Yes (0.25 eV) |

A conventional TES consists of a superconducting absorber whose lateral dimensions are larger than the detection wavelength. High efficiency optical coupling has been demonstrated by incorporating a dielectric stack anti-reflection layer.[7,8] An important issue for the proposed device is how to achieve efficient coupling of the incident signal to a nanoscale detector element. We propose to incorporate the device in a resonant near-IR antenna geometry, as discussed in section 4. Achieving efficient optical coupling will be important for the envisioned applications.

## 2. COUNT RATE

If the length $L$ of the Nb detector element is short compared to $\pi\sqrt{\tau_{e-ph}D} \approx 1\,\mu m$, the dominant cooling mechanism of the electron system is the outdiffusion of hot electrons into the leads.[13] Here $D$ is the diffusion constant, $\approx 1$ cm$^2$/s for Nb at 5 K, and $\tau_{e-ph}$ is the electron-phonon inelastic scattering time, $\approx 1$ ns at 5 K and proportional to $T^{-2}$.[14] In this case, the thermal time constant is $\tau_{th} = \tau_{diff} = L^2/(\pi^2 D)$. Thus, for $L = 200$ nm, $\tau_{diff} \approx 40$ ps. Such a fast response time requires readout with a 50 Ω microwave amplifier. As the device is also designed to have an impedance ≈ 50 Ω, the time constant will not be significantly affected by electrothermal feedback.

The thermal time constant is the time for the measured signal pulse to decrease to 1/e from its peak amplitude. A realistic definition of device reset is when the signal has decreased by 90% from its peak amplitude, which takes a time $\tau_{reset} \approx (2.3\tau_{th})$. This implies a maximum count rate of $1/(2\pi\tau_{reset}) \approx 1.7$ GHz for photons arriving at a regular time interval. If the photon arrival times are randomly distributed, then the effective maximum count rate is decreased by an amount that depends on the required detection efficiency.[2] This decrease in the count rate for randomly arriving photons can be minimized by splitting the signal between multiple detectors.

Another important timescale is the electron-electron inelastic scattering time $\tau_{e-e} \approx (10^8 R_{sq} T)^{-1}$, where $R_{sq}$ is the sheet resistance and $T$ is the temperature.[15] This is the timescale for the electron system in the Nb to reach a thermal distribution after absorbing a photon, and it determines the rise-time of the detection pulse. If $\tau_{e-e}$ becomes longer than $\tau_{diff}$, hot electrons will diffuse out of the Nb before the electron system reaches a thermal distribution. The initial energy sharing following photon absorption will be very rapid due to the high energy of the excitations, and will slow as the electron system approaches a thermal distribution at $T \gtrsim T_c$. In estimating $\tau_{e-e}$, we assume an average effective temperature of 10 K. With $R_{sq} = 25$ Ω, this gives $\tau_{e-e} \approx 40$ ps. The predicted maximum count rate for $L = 200$ nm is thus expected to be close to the maximum count rate achievable without experiencing degraded sensitivity due to the finite time required for the electron system to thermalize.

A faster count rate would be possible through the use of a different material. For example, thin-film niobium nitride (NbN) has $T_c \approx 10$ K and a resistivity that is approximately an order of magnitude larger than Nb. This results in $\tau_{e-e} \approx 2$ ps. The diffusion constant of NbN is $\approx 0.3$-$0.4$ cm$^2$/s,[16] and so for $L = 60$ nm, $\tau_{diff} \approx 10$ ps. Thus a NbN device should be capable of count rates well above 1 GHz even for randomly arriving photons. We note, however, that the use of a higher $T_c$ material will decrease the energy resolution, as discussed in the following section.

## 3. ENERGY RESOLUTION

The energy resolution $\delta E$ is the uncertainty in the measured photon energy, and quantifies the sensitivity of an energy-resolving detector. The energy resolution contains contributions from intrinsic device noise, as well as contributions from extrinsic noise sources such as the read-out amplifier. Intrinsic device noise includes statistical thermal fluctuations, which arise from the random exchange of energy between the device and its environment, as well as Johnson noise. All of these noise sources are uncorrelated, and hence each of their contributions to the total energy resolution add in quadrature.

In the absence of electrothermal feedback, the full-width at half-maximum (FWHM) energy resolution due to statistical thermal fluctuations is given by[17]

$$\delta E_{th} = 2\sqrt{2\ln 2}\sqrt{k_B T^2 C_e} \ . \tag{1}$$

In previous work on larger-area Nb microbridges, we determined $C_e$ from measurements of the thermal time constant and the thermal conductance.[18] In that work, we found $C_e = (23VT^3)$ JK$^{-4}$m$^{-3}$, where $V$ is the Nb volume and $T$ is the temperature. For $T = 5$ K and $V = $ (200 nm x 100 nm x 12 nm), we get $C_e = 6.9$ x $10^{-19}$ J/K. Using this value in equation (1), we find a FWHM energy resolution due to statistical thermal fluctuations of $\delta E_{th} = 0.22$ eV.

Amplifier noise can be expressed as an equivalent Johnson noise temperature. Both the Johnson noise of the detector and the amplifier noise temperature can then be converted to a noise equivalent power (NEP)

$$NEP = \frac{\sqrt{k_B T_N R}}{S_V} \tag{2}$$

where $R$ is the device resistance (assumed here to be 50 Ω), $S_V$ is the voltage responsivity, and $T_N$ is either the physical temperature of the Nb electron system (for Johnson noise) or the amplifier noise temperature. In the limit where the response frequency is small compared to the inverse thermal response time, the voltage responsivity is $S_V = I_{dc}(dR/dT)/G$, where $I_{dc}$ is the dc bias current, $G$ is the thermal conductance, and $dR/dT$ is the temperature-dependence of the device resistance. We estimate that $I_{dc} \sim 2$ μA, $G \approx 2$ x $10^{-9}$ W/K,[18] and $dR/dT \sim 100$ Ω/K, resulting in $S_V \sim 10^5$ V/W. The NEP is related to the FWHM energy resolution by[17]

$$\delta E = 2\sqrt{2\ln 2}\left(\int_0^\infty \frac{4}{NEP^2}df\right)^{-1/2} . \tag{3}$$

Equation (3) requires that one consider all contributions to the NEP. We can estimate the contribution to the energy resolution from Johnson noise or amplifier noise by converting the noise temperature to an NEP using equation (2) and then solving for the energy resolution in equation (3) using an upper limit of integration equal to the inverse of the detector time constant $\tau_{diff}^{-1} = (40 \text{ ps})^{-1}$. If we assume an amplifier noise temperature $T_N = 20$ K (a realistic value for a broadband cryogenic microwave amplifier[19]), we find $\delta E_{amp} = 0.11$ eV. The contribution from Johnson noise at 5 K is $\delta E_J = 0.05$ eV. Hence, considering the contributions from thermal fluctuation noise, amplifier noise, and Johnson noise, we get a total predicted FWHM energy resolution $\delta E = (\delta E_{th}^2 + \delta E_{amp}^2 + \delta E_J^2)^{1/2} = 0.25$ eV. We conclude that the detector sensitivity should be sufficient to resolve 1550 nm (0.80 eV) single photons.

## 4. OUTLOOK

Achieving high efficiency optical coupling is essential for many of the envisioned applications. At lower frequencies – from the microwave to the far-IR – efficient coupling to a sub-wavelength detector has been demonstrated by integrating the detector into a planar antenna geometry.[20] This approach does not extend in a straightforward way to the near-IR, as the inductive response of the charge carriers in a metal becomes significant as the frequency approaches the inverse of the scattering time. At these frequencies, one must account for the complex index of refraction of the metal. Previous work demonstrated improved near-IR coupling to a nanoscale detector using a gold dipole antenna, although this work did not determine the absolute coupling efficiency.[21]

We are presently performing numerical simulations of the coupling at 1550 nm free space wavelength to our Nb nano-TES. The results of these simulations will be presented in a future publication. These simulations require knowledge of the frequency-dependent complex index of refraction of the detector materials. The complex index of refraction of our thin-film Nb was determined via spectroscopic ellipsometry by J. A. Woollam Co. We are initially exploring two possible device geometries. The first is based on a gold dipole antenna, similar to ref. 21. The second is based on the length of the Nb detector element being chosen such that the Nb element itself acts as a half-wave dipole. In both approaches, the leads for biasing and readout are narrow and oriented perpendicular to the dipole element. The substrate material is C-plane sapphire. Coupling from the substrate-side of the detector takes advantage of the preferential coupling to the antenna from the higher dielectric medium.[20] A sapphire lens with a 1550 nm anti-reflection coating can be mounted on the reverse side of the substrate. By combining a resonant antenna geometry with an appropriate optical coupling structure, it is envisioned that efficient coupling at 1550 nm can be achieved. Meeting this challenge will be essential to demonstrating the utility of the proposed Nb nano-TES detector.

## ACKNOWLEDGEMENTS

This work was supported in part by NSF grants CHE-0911593 and DMR-0907082.